## ASTRONOMY

# Magnetic fields and relativistic electrons fill entire galaxy cluster

Andrea Botteon[1,2,3]*, Reinout J. van Weeren[1], Gianfranco Brunetti[3], Franco Vazza[2,3], Timothy W. Shimwell[1,4], Marcus Brüggen[5], Huub J. A. Röttgering[1], Francesco de Gasperin[3,5], Hiroki Akamatsu[6], Annalisa Bonafede[2,3], Rossella Cassano[3], Virginia Cuciti[3,5], Daniele Dallacasa[2,3], Gabriella Di Gennaro[5], Fabio Gastaldello[7]

The hot plasma within merging galaxy clusters is predicted to be filled with shocks and turbulence that may convert part of their kinetic energy into relativistic electrons and magnetic fields generating synchrotron radiation. Analyzing Low Frequency Array (LOFAR) observations of the galaxy cluster Abell 2255, we show evidence of radio synchrotron emission distributed over very large scales of at least 5 megaparsec. The pervasive radio emission witnesses that shocks and turbulence efficiently transfer kinetic energy into relativistic particles and magnetic fields in a region that extends up to the cluster outskirts. The strength of the emission requires a magnetic field energy density at least 100 times higher than expected from a simple compression of primordial fields, presumably implying that dynamo operates efficiently also in the cluster periphery. It also suggests that nonthermal components may contribute substantially to the pressure of the intracluster medium in the cluster periphery.

## INTRODUCTION

Diffuse synchrotron radio emission in the form of central radio halos and peripheral radio relics has been observed in a large fraction of massive and dynamically disturbed galaxy clusters (*1*). These sources indicate that turbulence and shocks driven by accretion processes are capable of dissipating part of their energy into the acceleration of relativistic particles and amplification of magnetic fields, at least in the central cubic megaparsec of the intracluster medium (ICM) (*2*). Cosmological simulations, however, predict that turbulence and shocks driven by accretion processes extend to much larger scales than those sampled by current observations of classic diffuse cluster radio sources (Fig. 1) (*3*), although it is still unclear whether these mechanisms can convert enough energy into relativistic components and magnetic fields to generate detectable radio emission.

To enter into this unexplored territory, we have observed the dynamically disturbed galaxy cluster Abell 2255 at redshift $z = 0.08$ with very deep 72 + 75–hour low-frequency observations at 49 and 145 MHz performed with the Low Frequency Array (LOFAR) low- and high-band antenna (LBA and HBA), respectively (see Materials and Methods for more details). For comparison, our 145-MHz observation has a noise ×60 lower and an angular resolution ×25 higher than that performed with the Westerbork Synthesis Radio Telescope at 150 MHz (*4*), which was the deepest observation available in the past decade of this cluster; deep observations at 49 MHz were never performed in the past on any object.

## RESULTS

Our images at 49 and 145 MHz in Fig. 2 show prominent diffuse radio emission that embeds the entire galaxy cluster, extending up to a projected linear size of 5 Mpc, well beyond the x-ray thermal emission from the ICM detected with the Röntgen Satellite (ROSAT) and engulfing all the previously observed radio sources that reside in the central region of the cluster: a halo, a relic, and a number of active galactic nuclei (AGN) with extended tails of magnetized relativistic plasma (*4–8*). These sources are labeled in Fig. 2. The very large-scale emission, which we refer to as "envelope," has low surface brightness [typical values of ~0.3 to 0.4 μJy/arcsec$^2$ at

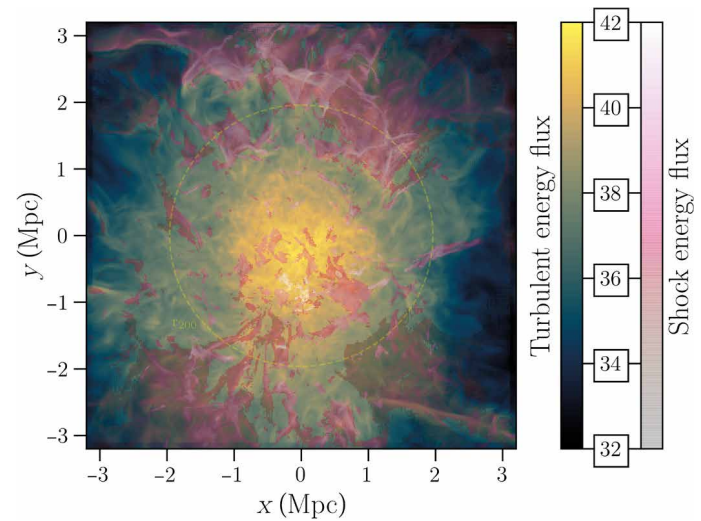

**Fig. 1. Turbulent and shock energy fluxes in a simulated galaxy cluster.** Colors represent the solenoidal turbulent energy flux and the shock (with Mach number > 2.3) energy flux in units of log erg s$^{-1}$ pixel$^{-1}$. Each pixel is 15.8 kpc by 15.8 kpc, and it gives the kinetic energy flux integrated along a column of 6.32 Mpc along the line of sight. The radius of the dashed yellow circle is 1.96 Mpc, corresponding to $r_{200}$.

[1]Leiden Observatory, Leiden University, PO Box 9513, RA, Leiden NL-2300, Netherlands. [2]Dipartimento di Fisica e Astronomia, Università di Bologna, via P. Gobetti 93/2, I-40129, Bologna, Italy. [3]INAF-IRA, via P. Gobetti 101, I-40129, Bologna, Italy. [4]ASTRON, Netherlands Institute for Radio Astronomy, Postbus 2, NL-7990 AA, Dwingeloo, Netherlands. [5]Hamburger Sternwarte, Universität Hamburg, Gojenbergsweg 112, D-21029, Hamburg, Germany. [6]SRON Netherlands Institute for Space Research, Niels Bohrweg 4, CA, Leiden NL-2333, Netherlands. [7]INAF-IASF, Milano, via A. Corti 12, Milano I-20133, Italy.
*Corresponding author. Email: botteon@strw.leidenuniv.nl

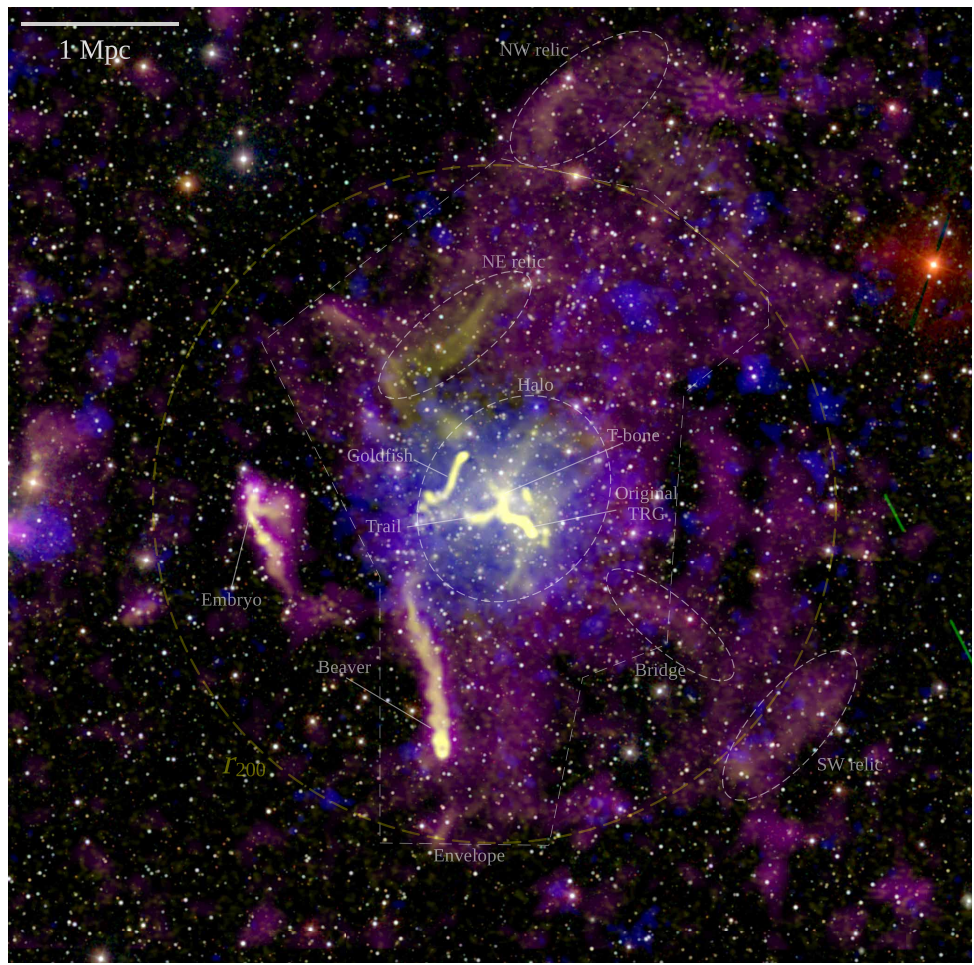

**Fig. 2. Composite image of the merging cluster Abell 2255.** Radio emission from LOFAR is shown in yellow (145 MHz, at high resolution) and magenta (49 MHz, at low resolution). X-ray emission from ROSAT in the 0.1- to 2.4-keV band is reported in blue. The background color image is from SDSS gri. The giant envelope of radio emission is labeled in white together with the other main radio sources discussed in the text, which names follow that used in previous studies (*5*, *7*, *8*). The radius of the dashed yellow circle is 2.03 Mpc, corresponding to $r_{200}$ (*53*). The field of view spans ∼1 deg² or roughly the size of four full moons.

145 MHz (1 Jy = $10^{-26}$ W m$^{-2}$ Hz$^{-1}$)] about 10 times lower than the radio halo at the cluster center. This envelope is clear in both the 49 and 145 MHz high-resolution images, but because of its extended nature, it is better recovered in the higher surface brightness sensitivity low-resolution images (60″) with discrete sources subtracted (Fig. 3 and fig. S1). The radio emission is brightest in the cluster center, where the deep high-resolution images reveal a large variety of substructures with filamentary nature (Fig. 4). Additional elongated and arc-shaped structures (relics) are clearly detected in the northwest and southwest peripheral regions, even beyond $r_{200}$, which marks the radius that encloses a mean overdensity of 200 with respect to the critical density of the universe at the cluster redshift (Fig. 2).

The observed synchrotron radiation requires 1- to 10-GeV relativistic electrons interacting with magnetic fields at about microgauss strength distributed on very large scale. While acceleration of nonthermal particles in collisionless astrophysical plasmas has been investigated in detail in supernova remnants, solar wind, and relativistic winds/jets (e.g., pulsar wind nebulae and AGN) (*9*), it is still unclear in galaxy clusters, where acceleration occurs in physical conditions that are unique in terms of spatial and temporal scales, collisional parameter, and plasma-β (*2*). Our observations now demonstrate that relativistic electrons are accelerated "in situ" and radiate in magnetic fields also in the very external regions of clusters, where the matter density falls to just ∼100 times the average value of the universe. These observations extend the possibility of exploring particle acceleration and magnetogenesis in regions where the ion-ion Coulomb collision time is 10 times larger than in cluster centers, about 30 million years, and the thermal ions and electrons Coulomb mean free path is very large, ∼25 kpc.

A map of the spectral index α (with flux density $S_\nu \propto \nu^{-\alpha}$, where ν is the frequency) for the diffuse emission in Abell 2255 is shown in Fig. 5. Excluding the contribution of the unsubtracted tailed AGN (see Materials and Methods), the spectral index values indicate the presence of both flat and steep spectrum emission (purple and yellow colors, respectively). The flattest spectrum emission is coincident with the filamentary and arc-shaped structures, while the steepest emission is coincident with the diffuse envelope. The histogram in fig. S2 shows the broad distribution of the spectral index values measured over the whole diffuse emission, ranging approximately from 0.5 to 2.5. The distribution is nonuniform and has likely different peaks, in line with a scenario where shocks and turbulence coexist and produce distinct spectral features. In particular, relativistic electrons are freshly accelerated near shock fronts and produce flat-spectrum emission, while particle acceleration by turbulence is a less efficient mechanism and thus tends to generate emission with steeper spectrum (2). The high-resolution images and spectral properties allowed us to disentangle shocks from turbulence as sources of particle acceleration (Fig. 4 and fig. S3). This picture, derived from the spectral index, is consistent with the polarized linear structures observed in the radio halo of Abell 2255 (10, 11), which are also suggestive of shock-powered emission located along the line of sight. Furthermore, a shock front associated with the bright and flat-spectrum radio relic located in the northeast, close to the cluster center, has been directly detected with x-ray observations (12).

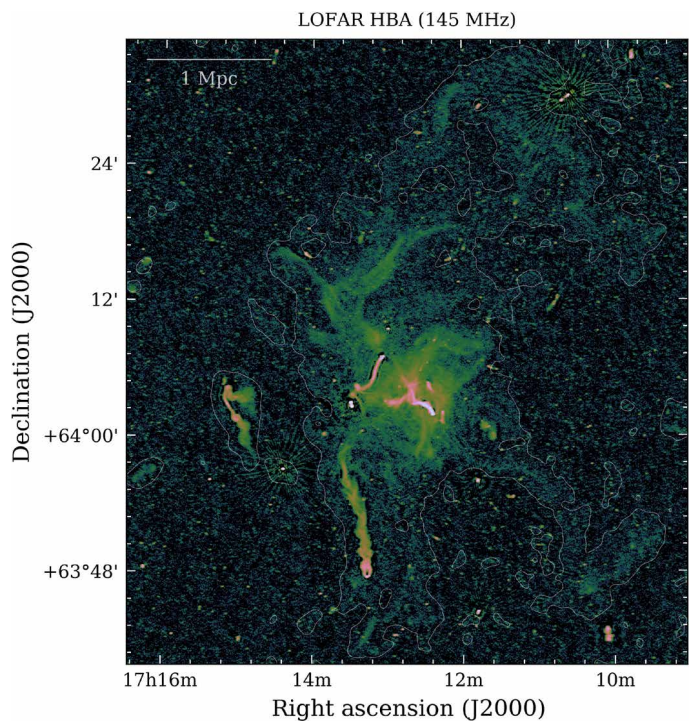

**Fig. 3. LOFAR 145-MHz radio image.** The image has an RMS noise of 43 μJy per beam and a resolution of 7.6″ × 4.7″. Colors represent intensity of radio emission. The white contours denote the 3σ level (with σ = 300 μJy per beam) from the LOFAR 145-MHz image at 60″ resolution after subtraction of discrete sources (except for the Beaver, Embryo, Goldfish, Original TRG, Trail, and T-bone radio galaxies; see Materials and Methods).

## DISCUSSION

The existence of detectable levels of radio emission at large distances from the cluster center allows us to provide a constraint on the efficiency of the particle acceleration and magnetic field amplification mechanisms in these poorly explored regions. From the observed synchrotron radiation, we can estimate the magnetic field ($B$) in the emitting volume (see the Supplementary Materials). Assuming that the relativistic plasma (electrons and $B$) has the minimum energy budget to generate the observed radiation and using a representative spectral index of $\alpha \sim 1.6$ (Fig. 5), we obtain $B_{min} \simeq 0.45(1 + k)^{0.222}(\gamma_{min}/1000)^{-0.409}$ μG, at a distance of ~2 Mpc, where $k$ is the ratio between the energy of protons and electrons and $\gamma_{min}$ is the minimum energy of relativistic electrons; a different $B$ implies a larger energy budget of the relativistic plasma. Specifically, at a distance of 2 Mpc, the electron thermal pressure is $\sim 1.3 \times 10^{-13}$ erg cm$^{-3}$ (13), which gives a ratio of relativistic to thermal energy density $\epsilon_{B+e}/\epsilon_{ICM} \sim 0.05 \frac{\Delta^{-2} + \mathcal{R}\Delta^{2.6}}{1 + \mathcal{R}}$, where $\Delta = B/B_{min}$ and the ratio between particles and magnetic field energy density is $\mathcal{R} = \frac{2}{1 + \alpha} \simeq 0.77$ (see the Supplementary Materials). This implies a magnetic field in the range $B \sim 0.1(\gamma_{min}/1000)^{-0.409}$ and $1.7(\gamma_{min}/1000)^{0.31}$ μG to prevent the energy content of the relativistic plasma from exceeding that of the thermal ICM (assuming that relativistic protons do not make a significant pressure contribution). We note that because we observe steep spectrum radiation at very low frequencies, $\gamma_{min}$ cannot be much larger than 1000, whereas a lower value of $\gamma_{min}$ would reduce the range of possible $B$ values.

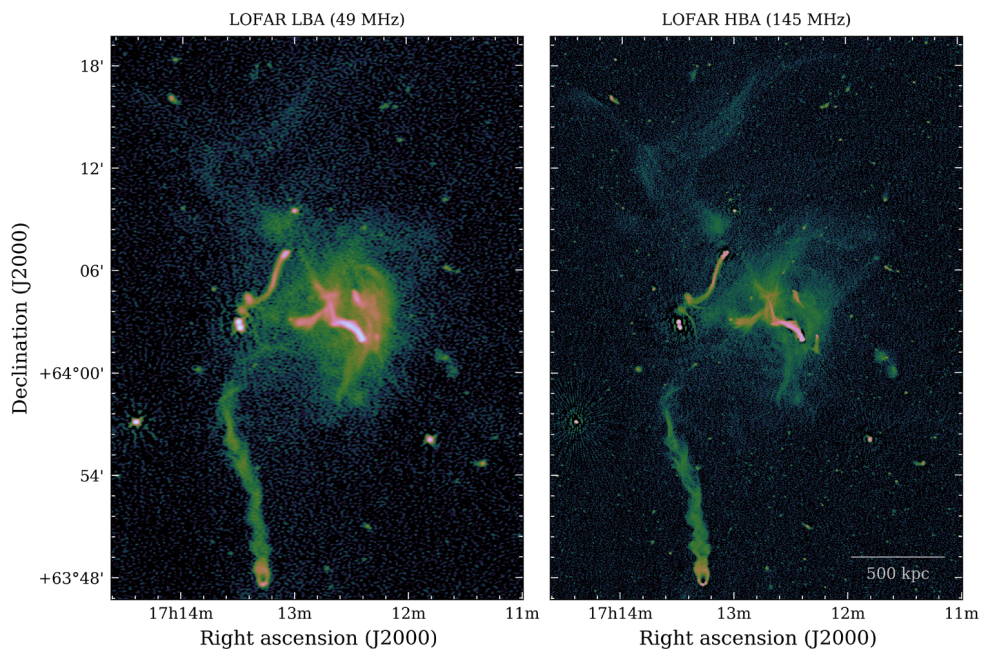

**Fig. 4. The highest-resolution LOFAR 49- and 145-MHz radio images obtained.** Colors represent intensity of radio emission. The 49-MHz image has an RMS noise of 0.73 mJy per beam and a resolution of 11.5″ × 8.2″. The 145-MHz image has an RMS noise of 55 μJy per beam and a resolution of 4.7″ × 3.5″.

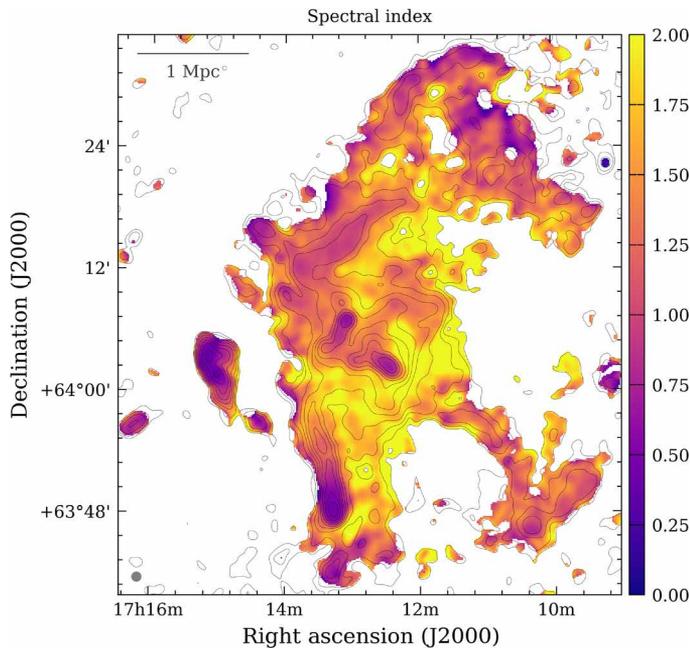

**Fig. 5. Spectral index map in the range of 49 to 145 MHz.** The map is obtained using LOFAR images at a common resolution of 60″ (corresponding to 91 kpc at the redshift of Abell 2255) with discrete sources subtracted (except for the Beaver, Embryo, Goldfish, Original TRG, Trail, and T-bone radio galaxies; see Materials and Methods). Pixels with surface brightness values below 2σ in the two images were blanked. Black contours represent the emission at 145 MHz and are spaced by a factor of 2 from 2σ, where σ = 300 μJy per beam. The beam is shown in the bottom left corner.

A possible explanation for the presence of extragalactic magnetic fields is that they have a primordial origin and formed in the early universe, before the epoch of recombination (14, 15). The lower bound of the magnetic field strength that we have estimated is ~250 times larger than the upper limit to the primordial magnetic field derived from the most updated modeling of the cosmic microwave background (16) and at least one order of magnitude larger than the value expected by the compression of such a primordial field, considering that we are observing regions with an overdensity of ~100. This suggests that the magnetic field should be considerably amplified by additional mechanisms operating in the cluster outskirts. It has been shown that the magnetic fields at about microgauss strength that are observed in the central regions of galaxy clusters can be produced by small-scale turbulent dynamo amplification of a seed field, provided that the effective Reynolds number in the ICM is large enough (17, 18). Under these conditions, magnetic field strengths of a fraction of microgauss are expected in cluster outskirts (19). We assume a scenario where turbulence amplifies the magnetic field via dynamo (20–22) and reaccelerates relativistic electrons. Our simulations (see Materials and Methods and the Supplementary Materials) predict that the energy flux of turbulence in the outer regions (1.6 to 2.2 Mpc) of an Abell 2255–like cluster, where we detect most of the external diffuse emission in Abell 2255, is $\sim 5-10 \times 10^{43}$ erg s$^{-1}$ Mpc$^{-3}$ (fig. S4). We assume that as a reference range of values to evaluate the fraction of turbulent energy flux that is channeled into nonthermal components (see the Supplementary Materials). Such energy flux, $F \sim \frac{1}{2}\rho \frac{\sigma_v^3}{\Lambda}$ [where ρ is the matter density and $\sigma_v$ is the turbulent root mean square (RMS) velocity at scale Λ], is converted into $B$ amplification, $B^2/8\pi \sim \eta_B F \tau_{eddy}$, where $\eta_B$ is an efficiency and $\tau_{eddy}$ is the eddy turnover time of turbulence, and into particle acceleration, which eventually generates synchrotron and inverse Compton emission, $L_{NT} \sim \eta_{acc} F V$, where $V$ is the emitting volume and $\eta_{acc}$ is an efficiency (see the Supplementary Materials). We find that to reproduce the giant envelope of emission observed in Abell 2255 (at 1.5 to 2 Mpc from the center), it is necessary that $\eta_B + \eta_{acc} \sim 0.05 - 0.1$, i.e., that 5 to 10% of the turbulent energy flux is channeled into nonthermal components (fig. S5). This suggests that the efficiency of amplification of magnetic fields in this tenuous and collisionless plasma is similar to that in magnetohydrodynamics (MHD) turbulence (20). Perturbations of the magnetic field created by collective effects in weakly collisional plasmas may decrease particle effective mean free path, making these plasmas "more collisional"; specifically, it has been shown that if the anisotropy relaxation rate due to the feedback of mirror and firehose instabilities in the high-β ICM is fast enough, the turbulent dynamo amplification may be similar to the collisional MHD case (23).

The combination of spectral index, surface brightness, and thermal pressure allows us to obtain further insights into the origin of the observed radio emission. One of the questions is whether relativistic particles are accelerated from the thermal matter or from a preexisting pool of suprathermal particles accumulated in the cluster but invisible in current observations. Our findings clearly rule out the former hypothesis. Even assuming minimum energy conditions of the relativistic plasma, we find that the spectrum of relativistic electrons cannot extend at energies below ~20 to 30 MeV (compared to few kilo–electron volts of the thermal matter); otherwise, the energy budget of suprathermal and relativistic electrons would immediately exceed that of the thermal matter (see the Supplementary Materials). Therefore, the nonthermal electrons generating the extended envelope of radio emission must be reaccelerated from an already existing suprathermal plasma. The cooling time of electrons in galaxy clusters is determined by Coulomb losses at lower energies and by radiative (synchrotron and inverse Compton) losses at higher energies (24). In cluster outskirts, the lifetime of electrons with energy ~100 MeV is of the order of the Hubble time (that is about 14 billion years), implying that relativistic electrons injected within the cluster volume can be efficiently accumulated, providing a reservoir of seed particles. In a few billion years, these seed particles can be spread and mixed on scales of several cubic megaparsecs by turbulent motions, losing memory of the spatial distribution of their origin. The main sources of seed electrons in the ICM are believed to be shocks occurring during the cluster's progressive assembly (25, 26). Further sources are individual galaxies that can inject high-energy particles through AGN outflows (27) or with phenomena related to star formation activity, such as stellar winds and supernova explosions (28, 29). Similarly, the outermost diffuse radio structures in Abell 2255 are likely the result of shock waves propagating in the cluster peripheral regions and crossing old populations of relativistic electrons. The shock origin of these sources is suggested by their arc-shaped morphology and the presence of a trailing edge. These outer relics are connected in projection to the central envelope of emission through low surface brightness bridges of emission and may produce detectable levels of radio emission via reacceleration mechanisms (*30*) or adiabatic compression processes (*31*).

The diffuse radio emission observed in Abell 2255 extends beyond the scales sampled by classical halos and relics, allowing us to probe ICM physics in an uncharted territory. This detection demonstrates the existence of cosmic rays and magnetic fields spread out to large cluster radii, indicating a considerable dynamical activity in these regions and an efficient conversion of the energy released during the large-scale structure formation process into nonthermal components. This confirms a critical piece of numerical simulations that until now could not be validated against any observations. These simulations also predict a substantial energy budget associated to turbulent motions, ~15 to 30% of the thermal energy (*32*–*34*). Our observations provide a first support to these predictions because we constrain an energy budget of nonthermal components (magnetic fields and relativistic particles) greater than or equal to several percent of the thermal budget, with the energy budget of nonthermal components being only a fraction of the kinetic energy budget (turbulence). Future observations will clarify how common this complex nonthermal phenomenology is among galaxy clusters and its interplay with cluster dynamics and cluster active galaxies.

## MATERIALS AND METHODS
### LOFAR observations, data reduction, and radio images
Abell 2255 was observed with LOFAR HBA stations as part of project LC12_027 for 75 hours and with LOFAR LBA as part of project LC15_024 for 72 hours. The HBA observations were divided into nine runs between 7 June and 15 November 2019 and followed the observing scheme of the LOFAR Two-metre Sky Survey [LoTSS (*35*, *36*)], namely, 8-hour observations on the target book-ended by two 10-min scans on the flux density calibrators (3C295 and 3C48) in the HBA_DUAL_INNER mode and with a frequency coverage of 120 to 168 MHz. All the data were passed through prefactor (https://github.com/lofar-astron/prefactor/) (*37*–*39*) to perform standard direction-independent calibration. Following the strategy used to process LoTSS-Deep Fields (*40*), we used ddf-pipeline (https://github.com/mhardcastle/ddf-pipeline) to build a sky model from a single 8-hour observation to calibrate the others and image the LOFAR field of view using killMS (*41*, *42*) and DDFacet (*43*).

The LBA observations were instead divided into 18 4-hour runs that were carried out between 10 December 2020 and 6 January 2021. Data were recorded in LBA_OUTER mode in the frequency range of 22 to 70 MHz, placing Abell 2255 and the flux density calibrators (3C295 and 3C380) into two different beams that were observed simultaneously for the whole duration of the experiment thanks to the multibeam capabilities of LOFAR. Demixing from Cassiopeia A and Cygnus A was used to subtract the signals of these two bright sources from the visibility data (*44*). Similar to the HBA observations, all the data were passed through prefactor before performing the final calibration using the Library for Low Frequencies (LiLF) framework (https://github.com/revoltek/LiLF) (*45*). As LiLF is not yet optimized for ultralow frequency observations, only data >30 MHz were kept after the prefactor step. The inspection of the direction-independent images produced by LiLF revealed that none of the 18 4-hour runs had to be discarded because of bad data quality (e.g., owing to severe ionospheric corruption). Thus, direction-dependent calibration and subsequent imaging were done with a slightly modified version of LiLF using the entire 72-hour dataset and the best model obtained from a single 4-hour observation, similarly to the strategy adopted for other LOFAR deep observations (*40*, *46*, *47*).

To further enhance the quality of the HBA and LBA images toward Abell 2255, we improved the accuracy of the calibration solutions with a postprocessing step. This is required, as the calibration solutions provided by ddf-pipeline are not optimized for the entire region covered by the cluster radio emission, which spans a sky area of about 1 deg². Therefore, we used the "extraction and self-calibration" method described in (*48*) to subtract all the sources located outside a square region of 1 deg² centered on Abell 2255 from the HBA and LBA visibility data to perform additional phase and phase+amplitude calibration loops on the smaller datasets. Each of the reprocessed datasets was thus jointly deconvolved with WSClean (*49*) to produce images with central frequencies of 145 MHz (HBA) and 49 MHz (LBA). The multiscale multifrequency deconvolution (*50*) option was enabled in WSClean during imaging to deeply deconvolve the extended and faint emission. During the imaging, we also adopted an inner $uv$ cut of 60λ, corresponding to an angular scale of ~57 arcmin, to reduce the sensitivity to the shortest baselines where the calibration is more challenging, and Briggs weighting (*51*) of robust = −0.5 unless stated otherwise.

To emphasize the cluster diffuse radio emission, we subtracted from the $uv$ plane the model of the discrete sources in the field and performed low-resolution imaging. The model used for the subtraction was obtained from high-resolution images allowing a minimum baseline of 1250λ to filter out the emission on physical scales larger than ~250 kpc at the cluster redshift. Before the subtraction, these models were visually inspected to ensure that any part of the cluster diffuse emission was not picked up in the process. The complex, bright, and extended radio galaxies termed Beaver, Embryo, Goldfish, Original TRG, Trail, and T-bone (*8*) and labeled in Fig. 2 were also not included in the models because of the impossibility of a reliable subtraction of their contribution. The RMS noise and resolution of the LOFAR images used in our analysis are summarized in table S1.

### Spectral index maps
We produced spectral index maps of Abell 2255 by combining 49- and 145-MHz images obtained at common resolutions of 12.5″ (~19 kpc at the cluster redshift; fig. S3) with discrete sources and at 60″ (~91 kpc at the cluster redshift; Fig. 5) with discrete sources subtracted out. Images were corrected for any position misalignment and regridded to the same pixelation. The spectral index was thus computed for each pixel where both images have a surface brightness above a given threshold (2σ or 3σ depending on the resolution) as

$$\alpha = \frac{\log\left(\frac{S_{145}}{S_{49}}\right)}{\log\left(\frac{49}{145}\right)} \quad (1)$$

where $S_{49}$ and $S_{145}$ are the flux density values at the corresponding frequencies. The error on the spectral index was obtained using the following formula

$$\Delta\alpha = \frac{1}{\ln\frac{49}{145}}\sqrt{\left(\frac{\Delta S_{49}}{S_{49}}\right)^2 + \left(\frac{\Delta S_{145}}{S_{145}}\right)^2} \quad (2)$$

where $\Delta S_{49}$ and $\Delta S_{145}$ are the uncertainties on $S_{49}$ and $S_{145}$, in which the flux scale error, which is 10% both at 49 and 145 MHz (*36*, *52*), and the image noise were added in quadrature.

We also made use of images obtained at common resolution of 35″ (~53 kpc at the cluster redshift) with discrete sources subtracted to produce another kind of spectral index map and study the spectral index distribution of the diffuse emission. In this case, spectral indexes were not computed on a pixel basis but within beam-independent square regions. This is required, as the beam in a radio image is covered by multiple pixels to satisfy the sampling theorem; thus, the neighboring pixels are not independent. The resolution of 35″ provides a good compromise to recover the extended cluster diffuse emission while allowing for discarding the contribution of the bright tailed radio galaxies that were not subtracted in the *uv* plane. Regions associated to these sources were blanked. The spectral index map and distribution of spectral index values are shown in fig. S2. The spectral index error maps are collected in fig. S6.

### Numerical simulations

In this work, we analyzed the $z = 0.02$ snapshot of a recent high-resolution cosmological simulation of a massive galaxy cluster resembling Abell 2255 (which is at $z = 0.08$), produced with ideal MHD grid code ENZO (enzo-project.org), taken from the suite presented in (*19*). From the cosmological point of view, there is no notable difference in the cluster structure between the two, so close epochs and the $z = 0.02$ snapshots of the simulations are those available at the closest redshift of Abell 2255. In particular, the simulated cluster (E18B) has a final total mass of $M_{200} = 8.65 \times 10^{14}$ $M_\odot$ within $r_{200} = 1.96$ Mpc, while Abell 2255 has $M_{200} = 10.33 \times 10^{14}$ $M_\odot$ and $r_{200} = 2.03$ Mpc (*53*).

The ENZO simulation includes eight levels of adaptive mesh refinement to increase the spatial and force resolution in most of the innermost cluster volume, reaching a maximum spatial resolution $\Delta x = 3.95$ kpc per cell across most of the volume within $r_{200}$. A uniform weak seed magnetic field of $B_0 = 10^{-4}$ µG (comoving) was assumed to be in place everywhere in the simulated volume at $z = 40$, mimicking a simple primordial magnetic field (*18*). However, the low redshift properties of the magnetic field in the cluster are fairly independent of the exact origin scenario because of the effect of the efficient small-scale dynamo amplification (*18*).

To compute the kinetic energy dissipation by turbulence and shocks ($F_t$ and $F_s$, respectively), we used the filtering technique described in (*3*), which allows us to robustly reconstruct the three-dimensional distribution of shock waves and turbulent gas motions through a combination of small spatial filtering techniques [see (*3*) for more details]. Once the $\sigma_v$ dispersion of the velocity field within a scale $\Lambda \approx 200$ kpc is measured via Helmoltz-Hodge decomposition, we computed the turbulent kinetic energy flux in each simulated cell as

$$F_t = \frac{\rho \sigma_v^3}{\Lambda} \Delta x^3 \quad (3)$$

where ρ is the gas mass density in the cell, Λ is the local turbulent scale measured by our algorithm, and $\Delta x$ is the cell size.

The kinetic energy flux across the shocked cells tagged in the computational domain by our algorithm provides a useful metric for energy available to the acceleration of cosmic rays by diffusive shock acceleration

$$F_s = \frac{\rho_u v_{shock}^3}{2} \Delta x^2 \quad (4)$$

where $v_{shock}$ is the shock velocity measured by our velocity jump–based algorithm, and $\rho_u$ is the upstream (i.e., pre-shock) gas density of shocked cells.

**SUPPLEMENTARY MATERIALS**

Supplementary material for this article is available at https://science.org/doi/10.1126/sciadv.abq7623

SCIENCE ADVANCES | RESEARCH ARTICLEBotteon et al., Sci. Adv. 8, eabq7623 (2022)    2 November 2022   8 of 8

**Acknowledgments:** LOFAR (54) is the Low-Frequency Array designed and constructed by ASTRON. It has observing, data processing, and data storage facilities in several countries, which are owned by various parties (each with their own funding sources), and are collectively operated by the ILT foundation under a joint scientific policy. The ILT resources have benefited from the following recent major funding sources: CNRS-INSU, Observatoire de Paris, and Université d'Orléans, France; BMBF, MIWF-NRW, and MPG, Germany; Science Foundation Ireland (SFI), Department of Business, Enterprise and Innovation (DBEI), Ireland; NWO, The Netherlands; the Science and Technology Facilities Council, UK; Ministry of Science and Higher Education, Poland; and Istituto Nazionale di Astrofisica (INAF), Italy. This research made use of the Dutch national e-infrastructure with support of the SURF Cooperative (e-infra 180169) and the LOFAR e-infra group, the LOFAR-IT computing infrastructure supported and operated by INAF, and the Physics Department of Turin University (under the agreement with Consorzio Interuniversitario per la Fisica Spaziale) at the C3S Supercomputing Centre, Italy. The Jülich LOFAR Long Term Archive and the German LOFAR network are both coordinated and operated by the Jülich Supercomputing Centre (JSC), and computing resources on the supercomputer JUWELS at JSC were provided by the Gauss Centre for Supercomputing e.V. (grant CHTB00) through the John von Neumann Institute for Computing (NIC). This research made use of the University of Hertfordshire High-Performance Computing Facility and the LOFAR-UK computing facility located at the University of Hertfordshire and supported by STFC (ST/P000096/1). The cosmological simulations were performed with the ENZO code (http://enzo-project.org), which is the product of a collaborative effort of scientists at many universities and national laboratories. We acknowledge the ENZO development group for providing extremely helpful and well-maintained online documentation and tutorials. The simulations were produced under project "radgalicm" at Jülich Supercomputing Centre (JFZ), with F.V. as principal investigator. We also acknowledge the usage of online storage tools provided by the INAF Astronomical Archive (IA2) initiative (www.ia2.inaf.it). This research made use of APLpy, an open-source plotting package for Python (55). **Funding:** A.Bot. and R.J.v.W. acknowledge support from the VIDI research programme with project number 639.042.729, which is financed by the Netherlands Organisation for Scientific Research (NWO). A.Bot. and A.Bon. acknowledge support from ERC Stg DRANOEL no. 714245. G.B., R.C., and F.G. acknowledge support from INAF mainstream project Galaxy Clusters Science with LOFAR 1.05.01.86.05. M.B. and F.d.G. acknowledge funding by the Deutsche Forschungsgemeinschaft (DFG, German Research Foundation) under Germany's Excellence Strategy (EXC 2121 Quantum Universe: 390833306). F.V. acknowledges financial support from the ERC Starting Grant "MAGCOW," no. 714196. A.Bon. acknowledges support from MIUR FARE grant "SMS." V.C. and G.D.G. acknowledge support from the Alexander von Humboldt Foundation. **Author contributions:** A.Bot. coordinated the research, wrote the manuscript, performed the data analysis, and led the LOFAR LBA proposal. R.J.v.W. helped with the data analysis, manuscript preparation, and led the LOFAR HBA proposal. G.B. carried out the theoretical analysis and helped with the manuscript preparation. F.V. performed the numerical simulations and helped with the manuscript preparation. T.W.S. helped with LOFAR HBA data analysis and the manuscript preparation. M.B. helped with the theoretical interpretation and the manuscript preparation. H.J.A.R. helped design the experiment and with the manuscript preparation. F.d.G. wrote the LOFAR LBA reduction software. H.A., A.Bon., R.C., V.C., D.D., G.D.G., and F.G. helped with the manuscript revision. **Competing interests:** The authors declare that they have no competing interests. **Data and materials availability:** All data needed to evaluate the conclusions in the paper are present in the paper and/or the Supplementary Materials. The observations are available in the LOFAR Long Term Archive (LTA; https://lta.lofar.eu/) under projects LC12_027 and LC15_024. The community-developed packages used to process the data are available in the following repositories: prefactor v3.0 (https://github.com/lofar-astron/prefactor/), ddf-pipeline v2.3 (https://github.com/mhardcastle/ddf-pipeline), and LiLF v1.0 (https://github.com/revoltek/LiLF). The electron thermal pressure profile of Abell 2255 was published (13) and is downloadable from the X-COP website (https://dominiqueeckert.wixsite.com/xcop/data).

Submitted 28 April 2022
Accepted 6 September 2022
Published 2 November 2022
10.1126/sciadv.abq7623